# The False Premises and Promises of Bitcoin

Brian P. Hanley

## Abstract

Designed to compete with fiat currencies, bitcoin proposes it is a crypto-currency alternative. Bitcoin makes a number of false claims, including: solving the double-spending problem is a good thing; bitcoin can be a reserve currency for banking; hoarding equals saving; and that we should believe bitcoin can expand by deflation to become a global transactional currency supply. Bitcoin's developers combine technical implementation proficiency with ignorance of currency and banking fundamentals.

**JEL:** E21, E22, E42, E51, G21, G29, G28

**Keywords:** Bitcoin, crypto-currency, cyber-currency, private currencies, alternative currencies

**Correspondence:** Brian P. Hanley, Butterfly Sciences, California, USA
Email: **brian.hanley@ieee.org**

The pre-publication comments of Geoffrey Gardiner and Edward Hugh on this paper are gratefully acknowledged.

# 1        Introduction[1]

Bitcoin is based on a paper by the pseudonymous Satoshi Nakamoto; it is a digital currency started in 2009 that creates unique, non-duplicable electronic tokens using software (dubbed mining) with an asymptotic limit of creation of 21 million tokens[1]. Every four years the number of bitcoins created is scheduled to be cut in half until 2140 when creation is supposed to go to zero. Mining is done by volunteers motivated by bitcoin rewards that operate servers running bitcoin software. The system operates by clearing transactions in a peer-to-peer decentralized system. Bitcoin provides for division of bitcoins into $10^8$ parts, dubbed satoshis.

The 21 million limit on the number of tokens is intended to create scarcity, in order to support pricing of those tokens in standard currencies. At time of writing, an estimated 11-12 million bitcoin tokens have been created, and an unknown number have been lost and cannot be remade. The tokens have neither intrinsic nor price supported valuation – their price floats on exchanges against world currencies. The ability to subdivide each bitcoin into 100 million satoshis is supposed to allow for expansion of the currency.

The bitcoin ecosystem includes electronic exchanges, and an implementation of privacy, such that it is possible to use bitcoins fairly anonymously without taking unusual measures[2].  Bitcoin provides an infrastructure for transfer of its tokens, and that infrastructure is integral with bitcoin's existence. Bitcoin can be exchanged for a fluctuating amount of various national currencies, with national borders the 'highwaymen' that users wish to avoid.

Some of the earliest adopters of bitcoin as payment have been those selling illegal goods and services[3, 4]. In addition to illegal drugs, prostitution, and contract killing, bitcoin money transfer systems can be used for trans-national asymmetric warfare, although there is no direct evidence that this has occurred.  In SEC v. Shavers, Mazzant ruled that bitcoins are money based on use as money and therefore investments made using bitcoin fall under regulation by the SEC[5].

---

[1] This paper was submitted to a couple of mainline economics journals. It was rejected because it was considered obvious, and hence, not novel enough research. I pointed out that this created a bizarre situation, in which one could write any number of pedestrian articles on bitcoin, and get published as novel. But because the errors of thought in the conception of bitcoin were so fundamental, pointing them out could not get published. Hence, this article in *ArXiv*. I do not have endless time, and my opinion that economic thinkers have an obligation to inform other fields of study is not shared by all. But this paper has been read over by experts in the field.

Today's bitcoin community tends to be insular, with active disinterest in entering the mainstream[6]. Bitcoin has attracted popular attention and some academic interest, including technical, legal, and the rare economic scholar. The claims of this cryptocurrency have virtually all been taken at face value, with little challenge to its fundamental design. However, by examining the premises of bitcoin, it becomes clear that virtually the entire enterprise is an intellectual house of cards.

The criticism herein is founded on fundamentals that have been almost completely forgone in the academic and popular record regarding bitcoin. One would hope that the errors discussed herein would be overwhelmingly obvious, but the publication record shows otherwise. Consequently, there is an "Emperor's New Clothes" cast to this critique, because bitcoin's errors are so basic.

## 1.1    Bitcoin representations

These premises, claims, and beliefs were derived from bitcoin FAQs[7, 8], forums, and articles[9, 10], and confirmed in conversations with proponents[2]. Most of these are direct quotes or nearly so.

1. *Bitcoin is one of the most important inventions in human history. It is the first time that the 'double spending problem' has been solved in software. Bitcoins can be put into a bank, and bitcoin loans can occur, just as with a fiat currency or gold standard currency bank.*

2. *Hoarding is another word for saving. Saving is much better than being in debt. Saving bitcoins leads to increased wealth as the bitcoin economy grows. Being in debt leads to interest payments and having less wealth.*

---

[2] Since initial publication, the author has discussed the contents of this paper with one of the primary cryptocurrency economists, Peter Surda. These representations were not a point of contention.

3. *Gold has the properties of being easily divisible and being of a limited supply [which] make it ideal as a currency. Bitcoins have the same properties of being easily divisible and of a limited supply.*

4. *Bitcoin proponents claim it can be expanded almost indefinitely by 'splitting' bitcoins into fractional coins. They claim that doing so functionally expands the supply of bitcoin.*

There are other representations by bitcoin proponents; however, it is not necessary to go through all of the ramifications that derive from these basic items. Seeing that list, most bankers will see serious problems on inspection, from fallacious reasoning to fundamental misconceptions.

## 1.2   Existing Bitcoin critiques and commentary

The European Central Bank, referencing the blog of Jon Matonis, a Forbes journalist on the board of the Bitcoin Foundation and a vocal proponent of bitcoin, voiced concerns that bitcoin has no intrinsic value and that bitcoin:

> *…fails to satisfy the 'Misean Regression Theorem', which explains that money becomes accepted not because of a government decree or social convention, but because it has its roots in a commodity expressing a certain purchasing power.* [11].

Krugman has criticized bitcoin because it incentivizes hoarding and creates deflation, but failed to note other problems[12]. Grinberg briefly touches on the possibility of bitcoin suffering a deflationary spiral, but otherwise discusses the ecosystem, technical, and legal problems; such as exchanges, potential failure of anonymity, denial of service attacks and violation of the stamp act[2]. Grinberg is an excellent reference for those interested in what bitcoin is and how it operates. Hoarding is tracked by Ron and Shamir[13], Mieklejohn, et al[14], as well as Sergio[15] without comment on its economic meaning. In addition, Mieklejohn, et al make an attempt to track circulation of bitcoins, claiming that roughly half circulate rapidly. However, since this occurs at gambling and trading sites, that activity does not represent buying and selling of goods and services.

Tyler and Moore show that patrons of bitcoin exchanges run significant risk of loss due to failure of the exchange[16].

Selgin is intrigued about a bitcoin type of crypto-currency within a fiat currency system as a way to provide a perfectly elastic currency supply that could be targeted by algorithm to various monetary schemes[17]. Selgin and Grinberg are both aware of issues inherent in an inelastic money supply. But the intractable nature of bitcoin's inelastic design is not connected by them with this issue.

A few focus on legality, providing introductory explanations of the technology of bitcoin[18, 19]. And recently:

> *California's Department of Financial Institutions has issued a cease and desist letter to the Bitcoin Foundation for "allegedly engaging in the business of money transmission without a license or proper authorization"* [20]

A precedent case for legality of bitcoin is the Liberty Dollar. However, this private currency was passed off as US currency, and contained considerably less value in silver than its face value indicated. Consequently, a significant part of that case was counterfeiting and fraud rather than stamp act violation[21]. Those features do not apply to bitcoin.

There are plenty of local or limited currencies which do not misrepresent themselves that are not prosecuted despite possibly violating the stamp act. Those currencies such as local scrip, grocery store coupons, and frequent flier miles, are redeemable in something that has a defined value in the fiat currency. Even Liberty Dollars had some intrinsic value in silver. Thus, bitcoin is unique in its pure market valuation.

Plassaras is concerned about the IMF being able to stabilize bitcoin, and states that bitcoin:

> *...poses a serious threat to the economic stability of the foreign currency exchange if it continues to grow in both value and usage. Any other digital currency that entered widespread use would pose similar problems.*[22]

This indicates that Plassaras believes that the valuation of bitcoins could reasonably be believed to be large enough that some fraction of them in the hands of one or two parties could launch an attack on the reserves of some national currency.

Some are concerned with the level of fraud and theft of bitcoins, translating into $5-$20 million in (nominal) criminal losses, together with $29 million seized by the FBI[23, 24], while others examine technical and organizational matters of bitcoin[25-28].

Jeong examines the anarchic political roots of bitcoin and cryptocurrency, as well as discussing fundamentals of the technology rather well[29]. She finds that the bitcoin cryptocurrency is part of the implementation of cypherpunk anti-government ideals along with Wikileaks, and has been identified by Julian Assange as part of the political effort of Wikileaks. She points out, correctly, that bitcoin is far less secure than commonly believed, and that bitcoin is a libertarian experiment. She identifies the irony of bitcoin's decentralized design being subsumed into dependency on a small number of exchanges and has a good discussion about bitcoin as an attempt at anarchic law. However, Jeong also fails to identify the fundamental problems of design within bitcoin I address.

Eyal and Sirer point out a serious technical vulnerability of bitcoin[30]. Bitcoin depends on the longest block-chain being the honest one. It has been understood from inception that this requires that the majority of nodes are honest. However, Eyal and Sirer describe a vulnerability that begins at 33% of computing resources. An implication is that a government (or wealthy private party) can take control of a cryptocurrency with this design (which is all cryptocurrencies now in existence) by applying superior computing resources. Even if the bitcoin algorithm is modified, it is evident that bitcoin will always be vulnerable to brute force application of sufficient computing resources to overwhelm the system.

Thus, it is apparent from examining the publication record, that bitcoin and its fundamentals are taken at face value with very few exceptions. It is also apparent that the fundamental errors in concept that will be shown date back to the origin of bitcoin, and misunderstandings about money and economics by the pseudonymous Satoshi Nakamoto[31].

## 2   Deconstruction of bitcoin

### 2.1   Bitcoin's purported capacity for expansion is not credible.

Bitcoin is currently designed to be divisible into units called satoshis that are 0.00000001 of a bitcoin[8]. That bitcoins are so divisible is supposed to mean that because 21 million bitcoins (the asymptotic limit) x 100,000,000 satoshis per bitcoin, is 2,100,000,000,000,000 (2.1 quadrillion) that bitcoin can be a viable global currency capable of supporting virtually any degree of expansion on the scale of nations.

Bitcoin proponents may take issue with the above statement because I did not subtract 1 from 2.1 quadrillion to symbolize staying below the asymptotic limit as Karpeles has done[8]. However, for these purposes, using a limit that is larger than the real one is just as meaningful.

Additionally, there are problems with Karpeles' calculation. Subtracting 1 from 2.1 quadrillion satoshis is not correct because bitcoins are mined in integer units, not satoshis, and cataloged losses of bitcoins have almost entirely been in whole units, at least 26,609 of them[24]. (In this use of the word "loss," a bitcoin loss is a documented removal from the system of a bitcoin entity that cannot be replaced.) So the minimum theoretical number of satoshis to subtract from 2.1 quadrillion is 100 million. Based on the 26,609 documented bitcoins lost, the actual upper limit is lower by at least that many bitcoins, which is 2.66 trillion satoshis. The true upper limit may be hundreds of thousands, perhaps millions fewer whole bitcoins, because people have reported losing digital wallets, and there is no visible difference between a lost bitcoin and a hoarded bitcoin[14]. In unauthenticated reports, people lost some large wallets in the early days when they didn't think bitcoins mattered and mining them was relatively quick.

Per figure 1, at the valuation of November 15, 2013, bitcoins sell on Mt.Gox for a nominal $430 each, with a wide range. The 2012 GDP of the United Kingdom in USD was $2.44 trillion[32]. If all bitcoins were available for use in commerce, then in order to support commerce equal to the U.K. alone, each bitcoin would have to appreciate 270 times from its current peak for a total of more than 2.3 million times its earliest valuation. Not gold, silver, diamonds, oil, beanie babies, nor any other valued commodity – not even tulip bulbs[33] has achieved that. All evidence available indicates that such a wild deflationary increase in valuation would not occur.

*Figure 1*: Bitcoin chart from Mt Gox exchange

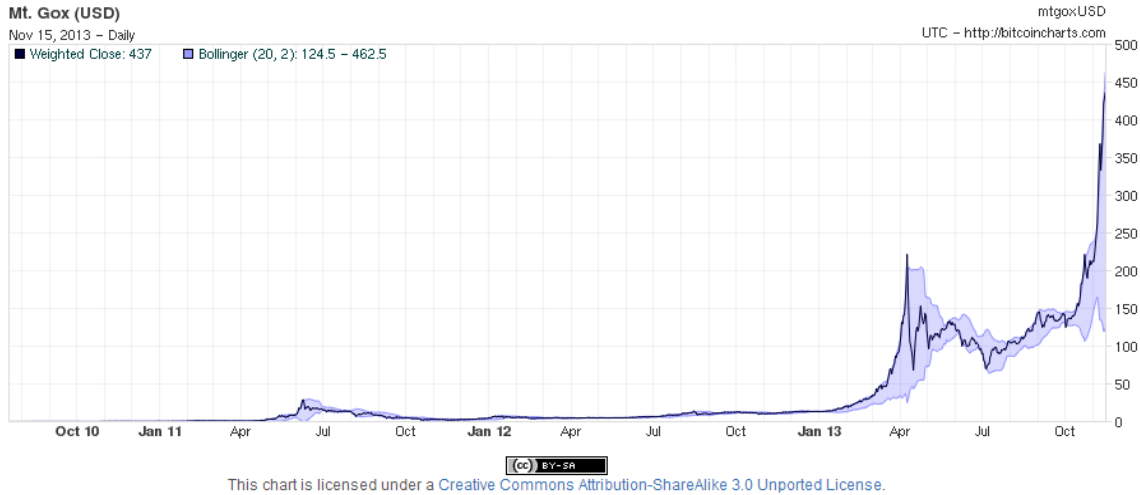

USD exchange price from October of 2010 through July 12, 2013[34]. Periods of increasing price are deflationary periods for bitcoin. Conversely, periods of decreasing price are inflationary periods for bitcoin.

In the real world, exchange rates for precious metals have not increased in price more than one or two orders of magnitude, even over periods of time like a century as shown for gold in figure 2.

*Figure 2*: USD price of gold (uncorrected for inflation) over 113 years

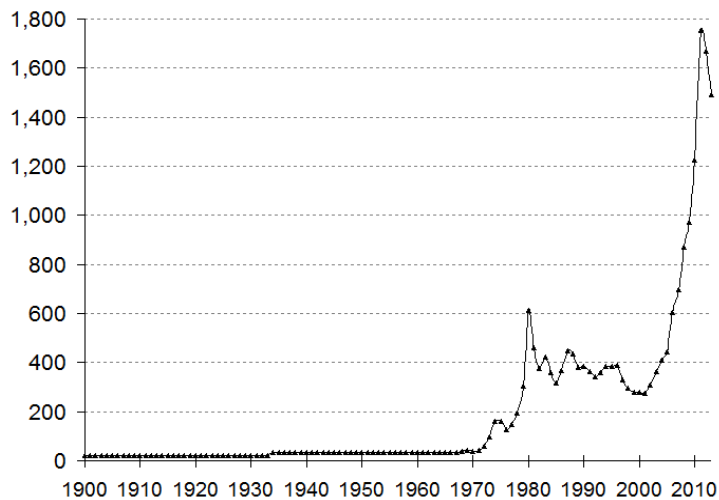

Uncorrected for inflation, the gold price per ounce over more than a century ranged from $20.67 to $1,791.75. The high is 86.68 times the low, which is less than 2 orders of magnitude. Beanie babies at their peak sold for roughly 2 orders of magnitude more than their nominal cost when the toys were first introduced. Corrected for inflation, gold increased by just 3.19 times over a century[35].

But let us forget about that and presume, for the sake of argument, that the USD valuation of each bitcoin rose to approximately $116,600 over the next 5 years as required to match a significant economy in the world. Generating a transactional economic value close to the UK's economy spending virtually all the bitcoins in existence each year would allow us to minimize the required rise in bitcoin valuation. Starting from the valuation of $430 per bitcoin would require bitcoin's valuation to multiply by 271 times over 5 years. That would be a 109% monthly compounded interest rate.

It is impossible to imagine that commercial trade transacted in bitcoins or centi-satoshis would be robust if the valuation were increasing at such rates. No rational player would use bitcoins for spending purposes. Certainly, there are irrational participants in every economy, but it is not in the least credible to believe that virtually every player, from the wealthiest to the poorest would spend large amounts of rapidly appreciating bitcoins every year. Nor is it credible to think that a fraction of players would spend so many bitcoins that their transaction volume would approach the necessary GDP through high velocity of money through the system. Without one or the other, the level of appreciation required to allow bitcoin to support an economy of a mid-size nation would be far higher. A higher rate of appreciation means an even greater incentive to hoard, which further decreases the credibility of bitcoin supporting actual commerce.

Consequently, it is impossible to imagine that the user base for bitcoins used in commerce could enlarge enough to drive such a valuation increase. The valuation of bitcoin will always be determined by speculation, not by utility for spending. It is believable that motivated transactors will continue make use of bitcoin as an alternative for a black and grey-market payment system, although regulators and law enforcement are making that more difficult. However, what will drive speculation is the creation of an enlarged, or simply wealthier, speculator pool.

Despite a report that the Cypriot financial crisis triggered the major rise in the price of bitcoin[36] the Cypriot crisis timeline does not line up well, although it may be possible that some account holders bought bitcoins. Lacking harder evidence, the entry of the Winklevoss twins appears most likely to have driven the most recent price rise – they claim to have acquired 1% of currently available bitcoins for their proposed ETF[20]. There is little evidence the speculator pool has enlarged much in numerical terms, but no statistics are published.

## 2.2   With bitcoin, reserve banking is impossible.

On a bitcoin "Myths" web page is a discussion of bitcoin and fractional reserve banking[8].  An anonymous blogger at Blogdial, cited by Matonis (who was in turn cited by the ECB) says:

> *When you have even a slight grasp of how data and computers work, and you understand that the double spending problem has been solved, your first reaction would be to gasp, as the enormity of what Bitcoin is dawns on you.*[37]

The double spending problem is the inability to transfer funds electronically without the use of a central clearinghouse that authorizes the transaction. Bitcoin has indeed solved that problem, (ignoring bitcoin's potential for takeover[30]) but a currency that *"solves the double spending problem"[10]* also ends banking as we know it.

The basis of banking is reserves[38, 39], (which reserves are now integrated around central bank money creation) and creation of new money through loans. That loan-created money is made of bookkeeping entries under the authority of banking regulators – it never exists as physical currency, and did not exist as physical currency in the heyday of precious metals[40]. Physical currency of any kind is a miniscule fraction of the money that exists.

The core of the issue is that since bitcoins are unique and cannot be duplicated, bitcoin can <u>only</u> exist as an electronic analog kind of physical coin. Ergo no money can be created by making a loan.

In the long past, enough gold or silver was, at least in principle, required to cover reserve requirements at a bank. The need for more gold to act as the core for banking reserves was once a major matter of concern for nations. Physical currency transactions in economies began to dwindle in the 14th century with the establishment of banks in Europe[41]. Gold and silver backed currency standards came and went versus fiat money in the 19<sup>th</sup> century. This continued until the formal ending of the gold standard in the USA in 1971, and in 2000 the formal end to the 40% backing of the Swiss Franc by gold.

Since all bitcoins are actual coin, the amount of bitcoin is limited, and bitcoins cannot be created on demand, it is impossible for bitcoins to be used to make loans since every loan would need to be made in actual bitcoins. To clarify this let's review a classical toy banking model based on 5% gold reserves as shown in figures 3 and 4.

*Figure 3*: Three iterations of loans in a 5% reserve banking system.

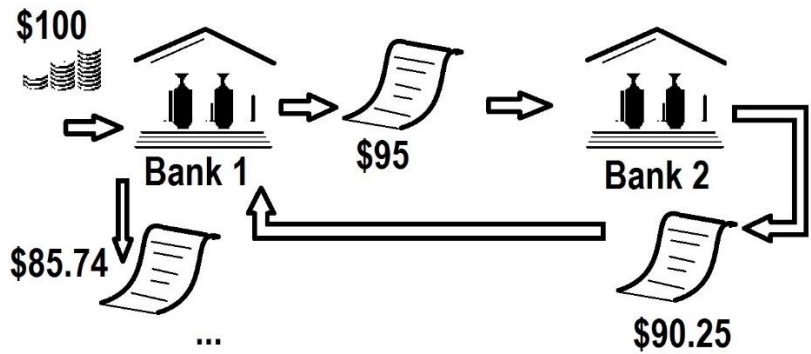

An initial hard currency (gold) deposit is entered into the books of Bank 1. Loan paper is created of 95% of the reserve. This is "virtual money" deposited into Bank 2. Bank 2 credits this virtual money and makes a new loan, of 95% loan which is deposited into Bank 1, and that in turn is accepted on Bank 1's books, a new loan is made, etc. The result is $95 + $90.25 + $85.74 = $189.49.  And that money creation can continue to the theoretical 1/r limit, where r is the reserve fraction required.

*Figure 4*: Physical coin system.

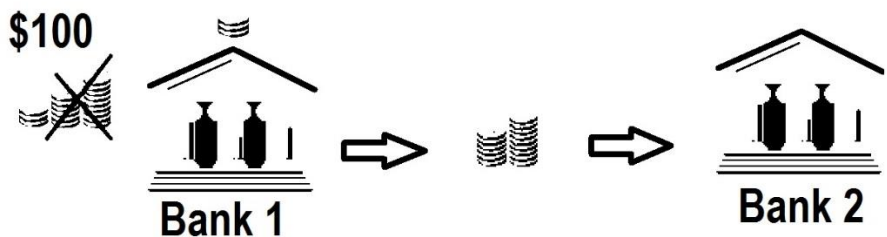

An initial gold deposit is placed in Bank 1 and logged into its books. Loan paper is created of 95% of the deposit. But this time the loan must be redeemed inside Bank 1 for the $95 in physical coin, and is carried out of bank 1 to deposit into Bank 2. When it is done, Bank 1 has $5 in coin and Bank 2 has $95 in coin. There is no change in the amount of money in the system, because no new money has been created by credit.

We have one of two choices here. We can allocate a new *virtual-bitcoin* to the depositor for 95% of the value of his deposit. Or, we can allocate the loan to as *virtual-bitcoin*, usable as if it were bitcoin, but not actually real bitcoin. *Virtual-bitcoin* is precisely the kind of money that bitcoin was designed to prevent, because bitcoin's designers did not think the problem through.

Figures 3 and 4 are schematics for a classical toy banking system based on a single gold deposit. In the real world, even for a gold-backed currency, things were more complex than shown. In the time of gold-backed currency, banks had capital reserves (today tier 1 and tier 2 capital, per Basel accords[39]) and those reserves were provided by the bank's partners or stockholders, not regular depositors. The diagrams here don't differentiate this.

Capital reserves ensured bankers had "skin in the game" that they would lose if their loans went bad. Their capital regulated how much deposited money could be loaned out. But records indicate that reserves in the old system varied widely. Even as long ago as the 1840's and before, in the heyday of gold-backed currency, a bank might operate at times with practically non-existent reserves, and this was fine for the economy[40]. Thus, the difference between gold-standard and fiat money of today is less clear from evidence than it is in theory.

*Figure 5*: Graphical representation of banking multiplier

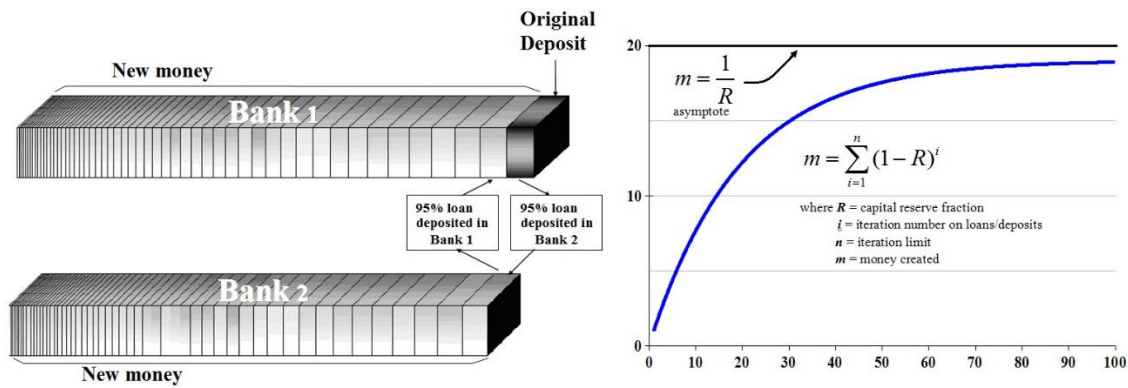

Each time a loan is made, it becomes a new deposit of a bank. The width of each brick in the above diagram is proportional to its size. The size of each loan declines because of reserve requirements. Equations of the banking multiplier are on the right. In practice, there are usually temporal limits to the banking multiplier, because originating a loan takes significant time. Also, loans are demand driven, which is why strategies like quantitative easing (QE) have trouble – QE is metaphorically pushing a rope.

Additionally, unlike the toy model, money from a loan would not necessarily come onto the books of a bank until it was spent. With gold and silver certificate paper notes, bank letters of credit, and bank cheques used to spend money, the net effect was similar to what is shown in figures 3, 4 and 5, but considerably messier. However, this classical toy model of banking has been good enough to educate beginning students for a long time, and is the basis for the mathematical derivation of the money multiplier asymptotic limit, so it is acceptable here.

Figures 3 and 4 make clear that creating loans based on bitcoin would require a new entity, the *virtual-bitcoin*, which would be backed by bitcoin, but not actually be bitcoin, just as gold-backed currency is backed by gold but not actually itself gold.

In this *virtual-bitcoin* scenario, bitcoin banks would keep bitcoin on reserve and redeem the *virtual-bitcoin* for real bitcoin in transfers, payments, etc. Such *virtual-bitcoins* would no longer be specific bitcoins that were deposited into an account, but instead be a *note* allowing the bearer the right to use it as if it were real bitcoin. This would correspond to a time in America many years ago when banks issued their own gold-backed currency, and the value of a bank's currency tended to vary with distance from the issuing bank.

No provision for *virtual-bitcoin* to exist in order to expand credit has been made in its design, and such ideas as paper currencies or accounting credits are anathema to the bitcoin community. The whole point of bitcoin is to force electronic transactions to only use these tokens that cannot be duplicated. To make *virtual-bitcoin* work would require a central clearinghouse to authorize the transactions, and then bitcoin would have come full circle – implementing the central clearinghouse accounting authority it was created to put an end to. Even if the objection of the bitcoin community to the idea of *virtual-bitcoin* could be overcome, it has other serious problems.

Primarily, why would a holder of a *virtual-bitcoin* note ever do anything except immediately present it for redemption in real bitcoin? We are not living in the naïve era of the Medici bankers, who could implement reserve banking without anyone being the wiser. Consequently the account holder would want to take possession of the underlying asset to prevent loss. I suppose some might prefer the *virtual-bitcoin* if enough interest was paid. But that would be certain to end in a bank run, and the result would look very similar to a Ponzi scheme.

Physical coin (gold, silver, etc.) is heavy, bulky and inconvenient. Bitcoin is not bulky – bitcoin has indeed solved that problem gold and silver have. All the bitcoins ever made could be held in a digital 'wallet' on a thumb drive. So the ancient motive of depositors to have a safe place to store their inconvenient, hard to safeguard money does not exist with bitcoin – except that bitcoin can be stolen[24]. But is the problem of potential theft large enough? And an even better question is, does risk of theft go up because of depositing bitcoins, or even trading them on an exchange? Evidence indicates it does[23, 24, 42].

The Bitcoinica web site created by a teenager was allegedly the site of a massive theft of bitcoins[43]. The operator of an entity in Texas, Bitcoin Savings and Trust, (initially named First Pirate Savings & Trust) has been arrested for defrauding depositors out of their bitcoins[5].

There is an entity that calls itself a bitcoin bank named Flexcoin[44]. But it does not offer loans; its FAQ claims that it only acts to facilitate transfers. It states that it charges a 1% fee for outbound transactions. It is not, in fact, a bank, and makes that clear on its web site. It is analogous to a hawala provider[45], although what anyone would want with a central clearinghouse that charges for transactions when the bitcoin infrastructure is free is beyond my ability to explain. That Flexcoin pays some kind of interest on accounts gives rise to serious questions, since the money transfer business model requires significant charges in order to make money. With the very low fee of 1%, one has to wonder how paying interest would be conducive to making money in that business.

The Flexcoin site has no contact information, no address, no phone number, not even an email. Attempting to find a contact by domain service lookup presents an obfuscated record through an entity in Paris.

It is hard to understand why anyone would use such 'banking' entities when it appears more work than using the existing bitcoin decentralized infrastructure. It is even harder to understand why someone would use them when it means turning over the tokens for a currency that is not legal tender in any nation to an entity that may be difficult to identify or locate in space-time.

## 2.3    Hoarding is different from saving.

*"Hoarding is another word for saving."*

Yes, money in a mattress is saved in the general sense of the word. But no interest can be had on that money. Healthy economies have some inflation, so hoarded money is worth less when when taken out of the mattress than when it went in. In addition, money in a mattress (or cupboard, jar, etc.) is vulnerable to being stolen. There has been notable theft and fraud of bitcoins[5, 24, 42, 43, 46]. And sometimes hoards, from pirate treasure to bitcoin wallets, are lost or forgotten[47]. Articles reference bitcoin 'brain wallets' that are dependent on a memorized passphrase for retrieval. If such a person suffers death, forgetfulness, or brain damage, their bitcoins will be lost forever to all.

In the world today, banking for most citizens is like the water a fish swims in. When money is saved it is typically deposited into a bank. This makes that money available for use in the wider economy through loans, since that deposit becomes usable by the bank (or credit union) to loan, which is, of course, part of how the banking system can multiply the quantity of money in the system as shown in figure 5. Making deposited money work through loans is how banks are able to pay interest on accounts.

Bitcoins, since they cannot be used in reserve banking, (see 2.2) can only be hoarded, spent, or lost, not saved in the usual sense it is thought of in the modern world.

## 2.4    Loans and interest payments on loans are the engine of wealth creation.

*"Being in debt leads to interest payments and having less wealth."* and *"Saving is much better than being in debt."*

Debt is the acquisition of money in the present in return for a promise to pay it back in the future. Of course it is possible for consumers to get into trouble by taking on more debt than can be paid back. This has long been a problem and always will be. The source of these memes is probably in a lack of discipline regarding taking on debt, and excessively lenient consumer credit leading to the perception by the over-indebted consumer that it is debt that is taking all their money. The indebted consumer in such a case does not connect to the macro-view that the debt extended to them was new money that entered into the economy in return for goods and services.

Business uses debt as a tool to create wealth. This is fundamental to our civilization. Classically, a business borrows money to buy raw materials and tools, and then creates a value added product that people will buy. It rarely is exactly that simple, but most businesses have a line of credit with a bank that they use to make payroll and pay other expenses during low points. A line of credit allows a business to continue operation in the sometimes very long lags between manufacturing, delivery, and getting paid. Truly, those who don't like debt don't like capitalism, because debt is another term for a loan, and loan credit is the bedrock of how capitalism works.

Businesses often extend credit, performing work and delivering a product or service before getting paid. The ordinary working person does exactly that over short periods of time by delivering a service before getting paid for it by their employer.

Wealth is goods and services in the "real economy". The financial economy is an abstract symbolic reflection of that[48]. We use money as a general medium of exchange, a store of value, etc. Creation of wealth in the real economy of goods and services requires debt. Yes, there are exceptions – some businesses operate successfully based on cash flow with no need for lines of credit; and some religious communes have operated successfully without any internal money, on the basis of mutual shared ideas, agreements, and rules for living[49]. However, even such communes used money externally, and their holdings were valued in external money of the larger society. In the social capital continuum[50], such communes are the high end. But in virtually all circumstances, debt is necessary to finance productive enterprises.

So the idea that debt leads to less wealth is backwards. In the larger economy, debt is the engine that leads to more wealth.

It is true that saving is a good thing when it supplies banks with more reserves so more loans can be made. The depositor has security for their money, receives payment that is generally above inflation, and makes capital available for loans. But the reason saving is good for society is because someone else is making use of the debt created by loans from the banking system and that creates more wealth in the real economy.

## 2.5    Saving bitcoins leads to increased (*personal*) wealth – but only when there is bitcoin deflation.

"~~Saving~~ [Hoarding] bitcoins leads to increased wealth as the bitcoin economy grows." This idea appears on its face to be self-evident. But in reality it is self-contradictory.

When bitcoins increase in value, they are deflating. Deflation is a characteristic of economic depression, not a growing economy, and it is the bitcoin economy that is supposed to grow. Deflation creates a liquidity trap for debtors [51] because their real interest rate is equal to the deflation rate times the formal interest rate of their loan.

Thus, if a month's deflation increases the valuation of a currency by 1%, and the monthly compounded interest rate is 0.5% (a yearly yield of 6.1%) then the true effective interest rate is 1.505%[3]. This looks small, but when compounded monthly, the yearly yield is 19.6%[4] per year, which is 13.5% above the formal 6.1% per year yield.

This can put debtors into a 'cash crunch' because they no longer can afford to pay their bills and service the loan. That means the bank will foreclose on the collateral for the loan, which is usually the assets of the business or individual. If it is a business, then the employees lose their jobs, suppliers don't get paid, and that business stops producing wealth in the real economy. Employees that lose their jobs can no longer afford to pay their bills and service their loans, and so other loans go under. That is the cycle that the Federal Reserve has been fighting with quantitative easing.

In addition, when money increases in value relative to goods and services of the real economy, then hoarding of money becomes a winning strategy. The higher the rate of increase in the value of money, the more effective is the hoarding (or miser) strategy. This may appear trivial to naive readers who might think that if the money is deposited into a bank that the bank can pay some level of interest, so it's all fine. But banks make their money on the spread between what they pay for money and what they receive in net return on loans. When the currency is increasing in value (deflating) what a bank is able to pay in interest can become less than zero. Less then zero is not typically an attractive interest rate to depositors.

When loans go bad in larger fractions than normal, the bank doesn't make as much money as it did and there are fewer creditworthy borrowers to pay the bank for loans. So depositors can't make much, and they may not be able to get their money back because the bank has too many bad loans. If depositors have amounts larger than the bank has in deposit insurance, then depositors can get caught in the downdraft. That sort of thing is what motivated people in the great depression to save their money in a mattress so they wouldn't lose it, and that crisis resulted in the FDIC.

As already seen, money that is hoarded is not productive money in the real economy because it is not invested and can't be made the basis for loans. Consequently, the economy has to suffer. Looking at the chart in figure 1, one can see a bubble occurred more than once that drove up the price of bitcoin, deflating the currency. Analyses of the bitcoin blockchain record show that hoarding is a serious issue[13-15].

---

[3] 1.01 x 1.005 = 1.01505

[4] $1.01505^{12} = 1.196$, or 19.6% interest.

Bitcoin proponents have answered criticisms about deflation with declarations that lack evidence. For instance:

> *"As deflationary forces may apply, economic factors such as hoarding are offset by human factors that may lessen the chances that a Deflationary spiral will occur."[8]*

In a time of rising prices, sellers would be interested in selling their bitcoins at the highest price they can. Conversely, buyers of bitcoins are less interested in taking them at the highest valuation because they may not be able to exchange their bitcoins for a less volatile currency before the bitcoin price drops, which tends to result in hoarding. This problem applies also to merchants. On highly volatile days, goods or services sold for payment in bitcoin could lose significant value in standard currency before exchanging them. So absent rigorous identification of what those claimed human factors are, together with a sensible model of human behavior during deflation, declarations such as the above are not credible.

Hoarding of a medium of exchange results in deflation (rising valuation) and a shrinking economy unless there are other currencies that predominate. In the case of bitcoin, those other currencies are the various fiat currencies of the world.

## 2.6    A bitcoin financial system is a losing zero-sum game for investors.

A system where the amount of money is fixed is a zero-sum game – for every winner, there must be a loser, because new money is not created that allows interest or investment return payouts. Bitcoin is designed to be a zero-sum game, and long before bitcoin creation is formally set to zero, the accidental loss of bitcoin wallets will match or surpass the creation rate. It is quite possible that this point has already been passed, but there is no way to monitor it because most bitcoins are hoarded, not used in commerce, and due to the distributed design, there is no visible difference between a hoarded bitcoin and a bitcoin that has been lost forever. Consequently, bitcoin is worse than a zero-sum game. It is a pulse game in which the bitcoin resource is injected and then slowly drawn down.

Without banking to make credit available, or without the ability to expand the money supply in concert with economic activity, interest payments in a zero-sum game can only cannibalize the money supply to pay winners. This forces a loss for every gain[52].

*Figure 6*: Interest rate, unfairness and investor return

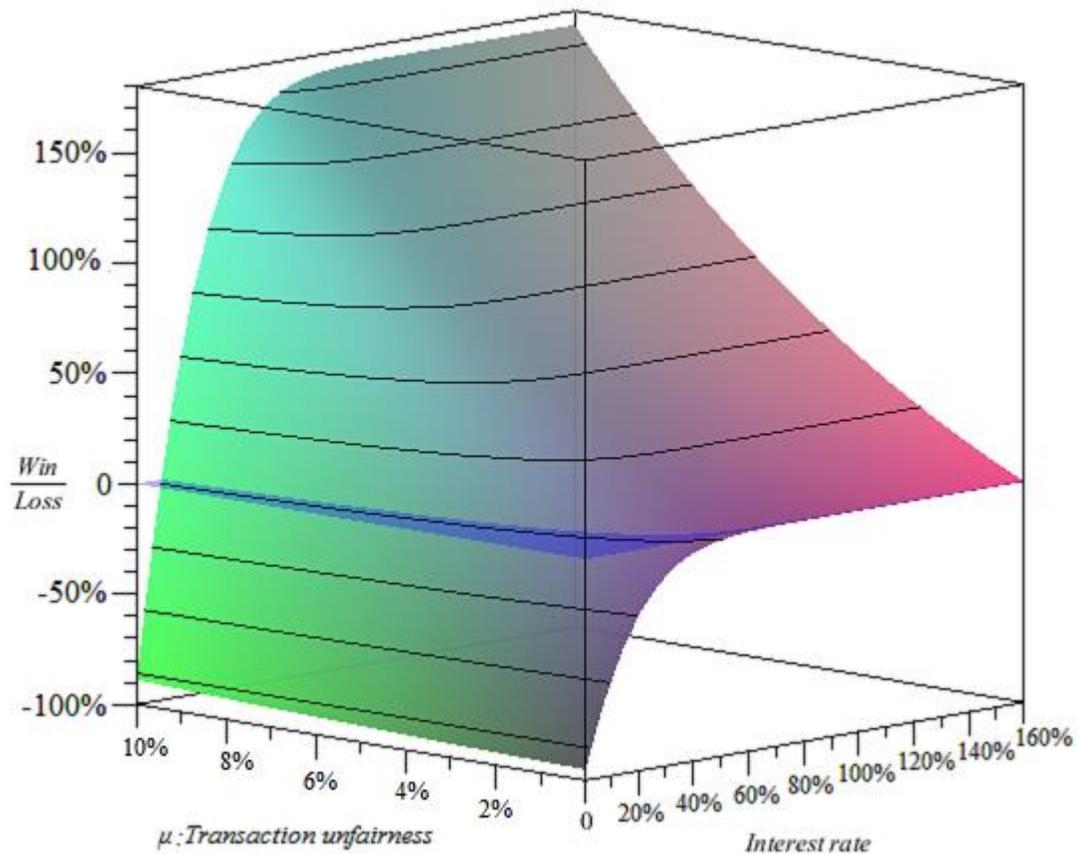

If the scales are unfairly tilted in favour of the investor class, the investor class can net a positive return, up to the limit of the money in the game. Within the investor class, probability dictates that some will be winners and others losers regardless of the interest rate, but the class as a whole will see these outcomes. Many will be surprised that to break even with a 5% advantage, the investor class requires an approximate 17% simple interest rate. [52]

A rational player rigs the game, or else charges outrageous amounts of interest/investment return. Even in a zero-sum game in which nobody understands what the rate of return requirements are, the system will evolve players who play the game according to winner's rules. We still have rules handed down from ancient times against usury and historical records of very high norms for interest rates in the past that indicate that ancient money-lenders evolved to conform to high interest rates.

This indicates that whatever bitcoin economy exists is dependent on the non-bitcoin economy for growth, because lending bitcoins in a pure bitcoin economy should have serious issues due to limitations on creation of money.

## 3    Conclusion

After having bitcoin explained to him, the most experienced banker I know said:

> *[bitcoin is] ...a very clever practical joke by someone who is having enormous fun exposing in the most sophisticated way imaginable the naivety of clever mathematicians, economists and/or rich speculators. ... or ... The cleverest con trick ever conceived, and probably one of the most rewarding.[53]*

My opinion is that bitcoin is most likely an accident born out of ignorance with some pecuniary interest thrown in. It should be obvious that even though bitcoin was created with built-in scarcity, every bitcoin in existence is itself newly created money. This is another irony of bitcoin – while bitcoin proponents decry the ability of governments to manufacture money, bitcoin is an entirely manufactured currency, which proponents intend to value in fiat currencies in order to profit. It should also be obvious that certain proponents have positioned themselves to make money by running exchanges and accumulating bitcoins. Those exchanges have no transparency, and the arms of regulators are only in the nascent phases of reaching bitcoin transactors. Bitcoin exchanges are positioned to trade on their own accounts in addition to charging fees.

An ECB publication states that bitcoin's theoretical roots are in Austrian economics[11]. Bitcoin corresponds with Austrian economic ideas in that bitcoin was intended to provide a monetary alternative that is beyond the reach of governments to regulate. Bitcoin has correspondence with libertarian ideas, which have some relationship with Austrian ideas. In the USA, my experience is that bitcoin proponents appear to have obtained their theory from science-fiction, radical libertarian popular literature, anti-government/anti-tax activism, and often from nothing that is apparent except their own thoughts[5].

---

[5] I certainly don't wish to discourage anyone from independent thinking. I merely wish to point out that there is a body of knowledge already developed which can improve understanding of economics, money and banking.

Bitcoin was developed by a motivated group of technologists who dreamed of creating a new currency that would cause fiat currencies to wither away. They believed that they had to do it with a distributed architecture that avoided a central clearing house in order to escape governmental control, an architecture that required that the tokens could not be duplicated. They wanted to do this because they believe that fiat currencies are the root of financial evil. They wanted to apply the Silicon Valley idea of 'disruptive technology' to the world economy.

There are precedents in history for successful disruption of finance. Hawala[54] type money transfer systems were disruptive. This invention of letters backed by the contents of a vault defended by a powerful clan was the first step toward changing the world. Those were hawala type practitioners, entities that took a commission in return for writing a letter to someone in another location so that the party using the service would not be required to move the material across regions where it could be stolen. That system established transactional convenience. Hawala was disruptive to those who made their living by robbery, and it enabled trade. But it wasn't yet banking.

The invention of banking was massively disruptive. It took power from royal families, making them beholden to bankers. The world we take for granted today in which a middle class life is presumed normal for most, a world where markers of health and wealth have vastly improved worldwide[55] could not exist without the banking revolution that put money creation into private hands. Banking made it possible to have competence win out over political favors to allocate capital. Banking created a revolution that made it feasible to extend credit at low rates of interest without necessarily having to rig the game – because the money system was no longer a zero-sum game. Very slowly, that revolution pushed the availability of credit downscale until today we take it for granted that virtually anyone in the developed world can get credit, and microloans are spreading all over the world. Ironically, it is the very ease of availability of that credit to consumers that fuels naive ideas of bitcoin proponents, such as that debt destroys wealth.

There may be a useful place for alternative forms of electronic money. However, an improvement requires study of money, financial institutions, finance history, and understanding of how and why our system works as it does today. At best, bitcoin is an unintentional throwback to pre-medieval finance.

As a currency to take over the world economy, or even a tiny part of it, bitcoin is not credible. Nor is there evidence in the history of commodity or currency valuations to suggest that the increase in price of bitcoins necessary to fulfill the dreams of proponents could reasonably be expected to happen. Similarly, there is no reason to think that a currency backed by nothing – a pure confidence currency – could overcome its hoarding and speculation problems and actually become an instrument used significantly for commerce.

I think that it would be helpful to put thought into developing systems that: A.) made credit more available to wealth-creating enterprises in the real economy; B.) improved evaluation of business ideas and startup teams so that loans could replace venture capital for many enterprises; and C.) improved allocation of capital within sectors so that capital does not repeatedly over-invest in the latest 'hot sector' thereby guaranteeing a lower average aggregate rate of return. In my view, those would be productive goals to work toward. Creation of a dysfunctional speculation vehicle is not a positive direction.

## 4   Afterword – Thoughts on how to disrupt finance in a positive way

This paper was primarily written in hopes that those involved with bitcoin can be reached and shown certain errors. The bitcoin community comes from an industrial sector that pays homage to disruption and Schumpeterian entrepreneurial spirit. What that community needs to understand is that banking is the original "disruptive technology". Banking changed the world. If you look around you, most of what you look at, from the desks, tables, computers and walls, to the clothes on your back, exists because banking made it possible. Banking is still evolving. In this afterword, I will provide my thoughts on how that spirit can be applied to finance, but in a positive way.

Within enterprise investment, I have observed first-hand that the people who make decisions to invest or loan money are often as ignorant of the area they invest in as bitcoin developers are of banking. Likewise, those who understand a technology are often equally ignorant of finance and business organization. These are problems that need to be solved somehow in order to serve the public good.

Perhaps a public market for underwriting of credit default swaps (CDS's) on real-goods/services enterprise investment could help to crowd-source investment decision-making. The software industry likes to think of itself as the originators of crowdsourcing, but stock and bond markets are the original crowdsourced decisions. When looking at what AIG did wrong, it was not writing CDS contracts, per se, that was the problem, it was a risk model that didn't coincide with reality. Perhaps that invalid model was deliberate, since the public would be on the hook to pay, and executives made bonuses while the music played, but that is speculation, and a separate problem.

In my opinion, CDS contracts are inappropriate for relatively fixed assets like real estate, because new value creation is almost entirely in the initial development. The temptation to abuse CDS contracts is great. However, they could be a good thing for enterprises that are creating real value. Using CDS contracts with the current Federal Reserve rules allows for very high growth – if it is warranted[56]. It was precisely that explosive growth that generated the housing bubble. In itself, rapid money creation isn't wrong. However, it needs to reflect something real, without overheating.

It used to be that an investment bank was a special kind of entity that was allowed to risk the money of its participants in investments. This made a lot of sense. If you can invest your money as a bank, then you can improve your returns.

Over time, investment banks like Goldman Sachs got more and more freedom, until they are no longer anything like what was originally intended. I think that we need something like an investment bank – perhaps we could call it a "Venture Bank" and allow it special privileges similar to the old rules for investment banks, but only if it will invest directly in real enterprises directly producing value.

Another idea that might help would be to create a banking infrastructure entity that could allow individuals or groups to manage their own money as an investment bank. Some floor of assets is needed, but aside from that? Why not give those people access to the Central Banks, just like the majors have? In this internet age, it could be done. It would allow people to learn how banks really work, and understand better where money comes from. It could, perhaps, even the playing field, vis-à-vis the giant banks. And it just might be possible to implement within the political system we have.

All of those readers who are newly minted millionaires and billionaires, think about it. Why shouldn't you be able to maximize the utility of your money by operating your investments as a bank? Back in the 19th Century, bankers like J.P. Morgan rocked the world. Morgan backed Tesla (the man, not the car company) and much else. Bankers with vision built things. But the gold standard meant the world banking system would periodically bump up against the limits of money creation. (Look at figure 5, right. When you are on the steep side of that curve, money is easy. The closer it gets to the ceiling, the tighter money gets.)

The gold standard problem is what the Federal Reserve/Central Bank system fixed. Yes, I am aware that many involved with bitcoin believe that "The Fed" is the root of all evils, etc. I am all too aware of the Tea Party movement's Alice-In-Wonderland views. "The Fed" and Central Banking is not "the problem". But virtually everything else is. The Fed and Central Banks around the world are the best functioning parts of our finance system today.

I have also seen first-hand that venture capitalists are lemmings. This isn't a new observation. I got my first invitation to lecture at the Leavey School MBA program after an argument with a professor about this issue. You see, what the lemming method accomplishes is to guarantee lower rates of return for venture capital. If a sector is over-invested, then it's obvious that many are going to fail – not because of incompetence – it's pure numbers. However, the lemming methodology is the inevitable outcome of the finance sector's inability to evaluate well what they are investing in. What else can they do? Certainly, deciding when a sector is over-invested, or even what the boundaries of a sector are, is a fuzzy problem – it's art, not science. But there are points when everybody with their hands in knows that there are too many.

So, to those interested in creative disruption, I would recommend:

- Investment/venture banking for the masses, or something like it.
- Venture banking to bring back what investment banks once were.
- Open-outcry exchange for all CDS contracts.
- Attempting to develop CDS type contracts on investments in startup and existing enterprises.

- Improving the connection between startup tech/ideas, business organization and investment.

That could be disruptive in a positive way.